# Precipitation of amorphous ferromagnetic semiconductor phase in epitaxially grown Mn-doped Ge thin films

Satoshi Sugahara[1,2,*], Kok Leong Lee[1], Shinsuke Yada[1], and Masaaki Tanaka[1,3]

[1]*Department of Electronic Engineering, The University of Tokyo, 7-3-1 Hongo, Bunkyo-ku, Tokyo 113-8656, Japan.*
[2]*PRESTO, Japan Science and Technology Agency, 4-1-8 Honcho, Kawaguchi, Saitama 332-0012, Japan.*
[3]*SORST, Japan Science and Technology Agency, 4-1-8 Honcho, Kawaguchi, Saitama 332-0012, Japan.*
*Present address: *Department of Frontier Informatics, The University of Tokyo, 5-1-5 Kashiwanoha, Kashiwa-shi, Chiba 277-8561, Japan*
Email: sugahara@cryst.t.u-tokyo.ac.jp

We investigated the origin of ferromagnetism in epitaxially grown Mn-doped Ge thin films. Using low-temperature molecular beam epitaxy, Mn-doped Ge films were successfully grown without precipitation of ferromagnetic Ge-Mn intermetallic compounds, such as $Mn_5Ge_3$.  Magnetic circular dichroism measurements revealed that the epitaxially grown Mn-doped Ge films exhibited clear ferromagnetic behavior, but the Zeeman splitting observed at the critical points was not induced by the s,p-d exchange interactions.  High-resolution transmission electron microscopy and energy dispersive X-ray spectroscopy analyses show phase separation of amorphous $Ge_{1-x}Mn_x$ clusters with high Mn content from a Mn-free monocrystalline Ge matrix.  Since amorphous $Ge_{1-x}Mn_x$ was characterized as a homogeneous ferromagnetic semiconductor, the precipitation of the amorphous $Ge_{1-x}Mn_x$ clusters is the origin of the ferromagnetic semiconductor behavior of the epitaxially grown Mn-doped Ge films.

Over recent years, ferromagnetic semiconductors (FSs), in which the lattice sites of the host semiconductor (SC) are partially occupied by magnetic ions, have attracted considerable attention for the research field of spintronics, since their small lattice mismatch and small conductivity mismatch with SCs and their extremely high spin polarization or half-metallicity are expected to be useful for ferromagnet/semiconductor hybrid devices [1-4]. Furthermore, the carrier-mediated ferromagnetism in FSs gives an additional functionality for manipulating spin degrees of freedom [5,6]. In particular, Si- and Ge-based FSs, which are expected to be compatible with advanced Si technology, have a potential impact on a new development of integrated electronics employing spin degrees of freedom, since they can play significant roles in recently proposed 'spin transistors' that are functional building blocks for such integrated circuits [7-9]. It should be emphasized that Ge and also SiGe are now well recognized to be key materials for advanced Si transistors with high performance [10-12].

After the first report of the epitaxial growth of ferromagnetic $Ge_{1-x}Mn_x$ films and their field-effect control of the ferromagnetism by Park *et al* [13], many experimental and theoretical investigations have been performed for Ge-based FSs [14-19]. Ferromagnetic $Ge_{1-x}Mn_x$ film/bulk samples were successfully grown/synthesized by several techniques, and high Curie temperatures close to room temperature were also reported. These results are likely to be supported by theoretical predictions for $Ge_{1-x}Mn_x$. Nevertheless, owing to the lack of experimental evidence for the s,p-d exchange interactions in $Ge_{1-x}Mn_x$ (which are essential for the ferromagnetism in FSs), controversy and reexaminations arose over the origin of the ferromagnetism in $Ge_{1-x}Mn_x$ [20-22]. Although the s,p-d exchange interactions in FS modify the band structure of the host SC (e.g., spin splitting of the band structure), the basic character of the band structure of FS well reflects that of the host SC, and thus the spectroscopic features of FS are strongly correlated with those of the host SC. Therefore, the spectroscopic characterization of the band structure of FS, including the Zeeman splitting at critical points induced by the s,p-d exchange interactions, is important to clarify the origin of the ferromagnetism in FS. However, this has not been explored for $Ge_{1-x}Mn_x$ so far .

In this paper, we investigated the origin of the ferromagnetism in epitaxially grown Mn-doped Ge films, based on the microscopic structure analyses using high resolution transmission electron microscopy (TEM) combined with energy dispersive X-ray spectroscopy (EDX) and on the direct band structure characterization using magnetic circular dichroism (MCD). It is shown that the precipitation of amorphous $Ge_{1-x}Mn_x$ nanoclusters, which were found to be a ferromagnetic semiconductor phase, is the origin of the ferromagnetism in epitaxially grown Mn-doped Ge films.

Mn-doped Ge films were grown on Ge(001) substrates by low-temperature molecular beam epitaxy (MBE). The Ge substrates used in the experiments were of commercially available undoped Ge(001) wafers with the resistivity of > ~40 $\Omega$cm. After a Ge(001) substrate was cleaned by acetone and deionized water and then the native oxide layer of the Ge substrate was etched by HF solution, the substrate was transferred into the ultra-high vacuum MBE chamber using an oil-free load-lock system in order to avoid any carbon-related contamination. After thermal treatment of the substrate in a ultra-high vacuum, a 15-nm-thick Ge buffer layer was grown at a substrate temperature $T_S$ of 100°C and successively a 30-nm-thick Mn-doped Ge film was grown at $T_S$=100°C. The Mn concentration was controlled by the intensity ratio of the Mn and Ge fluxes ($F_{Mn}/F_{Ge}$). The growth rate was 30nm/h for all samples. The time evolution of the growth mode and crystallinity during the epitaxial growth were observed by *in-situ* reflection-high-electron-energy-diffraction (RHEED). The nominal Mn concentration $x_N$ was measured by secondary ion mass spectroscopy (SIMS) with $XCs^+$ technique (that is effective to exclude the matrix effect on SIMS measurements). The microscopic crystal structure and chemical composition of the films were analyzed by high-resolution TEM and local area EDX. MCD was used for the evaluation of the s,p-d exchange interactions. MCD is defined by the difference in optical reflectance (absorption) between right- and left-circular polarized lights. It is proportional to the Zeeman splitting of the energy band of a FS induced by the s,p-d exchange interactions, and MCD is also proportional to the magnetization of the material. The importance and features of MCD characterizations on FS were pointed out by a couple of papers, in judging whether the material is an intrinsic FS or not [23-27].

After the epitaxial growth of a Ge buffer layer, the RHEED pattern of the surface exhibited clear streaky 2×2 reconstruction. When an Mn-doped Ge film with $x_N = 6$ % was grown on the buffer layer, a transition from the streaky 2×2 reconstruction to 1×1 streaks was observed at the initial stage of the epitaxial growth. The intensity of the diffraction pattern was slightly weakened with increasing the thickness of the Mn-doped Ge film, and then elongated spots were superimposed on the streaks, indicating an intermediate growth mode between layer-by-layer growth and three-dimensional growth. This growth mode was stably maintained up to the film thickness of 30 nm (that is the maximum thickness examined here). The qualitatively same growth kinetics was obtained for Mn-doped Ge films with $x = 1-6$ %. When the Mn concentration increased to $x_N = 12$ %, the RHEED pattern changed quickly from the 2×2 streaks to 1×1-elongated spots at the initial stage of the epitaxial growth, i.e., the three-dimensional growth. This surface roughening might be caused by the formation of clusters of Ge-Mn intermetallic compounds such as $Mn_5Ge_3$. This scenario is consistent with the fact that the Mn-doped Ge film with $x_N = 12$ % showed a Curie temperature $T_C$ of ~290 K which is close to that of $Mn_5Ge_3$ ($T_C = 296$ K) and its temperature-dependent magnetization (*M-T*) curve exhibited a bimodal shape (that implies the existence of two critical temperatures for the ferromagnetic ordering), indicating the precipitation of $Mn_5Ge_3$ clusters. Recently, it is shown that high Curie temperatures close to room temperature observed in Mn-doped Ge films is caused by the precipitation of Ge-Mn intermetallic compounds. [20, 21]

Figure 1(a) shows a cross-sectional TEM image of an epitaxially grown Mn-doped Ge film with $x_N = 6$ %, where the scanning area was 210 nm in width. Although many bright contrast regions with a columnar shape were observed in the epitaxially grown film as shown in Fig. 1(a), a high-resolution TEM lattice image (not shown) of the sample showed a complete diamond lattice structure alone without any cluster and grain boundary. However, we cannot exclude the possibility that such a lattice image appears for the following reason: When *amorphous clusters* with the diameter less than the thickness of TEM specimens (with respect to the projection direction) were embedded in the film, the matrix surrounding the amorphous clusters shows its diamond-type lattice image on the cluster regions without inducing any other lattice image or any moiré pattern. Figure 1(b) shows a cross-sectional TEM lattice image of the same Mn-doped Ge film, after the thickness of the TEM specimen was thinned by re-milling. Phase separation of clusters and a crystalline matrix was clearly observed in the epitaxial film. Note that the lattice image with a dark contrast was observed at the bottom of the cluster, which is caused by the residual matrix on (or beneath) the cluster. The transmission electron diffraction (TED) pattern of the cluster region was a halo pattern without diffraction of any other lattice structure, indicating an amorphous phase (Note that the diffraction from the matrix surrounding the cluster was superimposed on the halo pattern, owing to the very narrow width (~5 nm) of the cluster, i.e., the TED pattern includes the information of the matrix surrounding the amorphous cluster). The TED pattern of the matrix region of the Mn-doped Ge layer was exactly the same as that of the Ge buffer layer and no lattice distortion was observed. Thus, the matrix has the diamond lattice structure of bulk Ge. Local area EDX measurements showed that the Mn concentration $x$ of the cluster shown in Fig. 1(b) was 18 %, and Mn concentrations ranging from 10 to 20% were detected from other amorphous clusters formed in the epitaxial film. However, the Mn concentration of the matrix region was under the detection limit of our EDX measurements. Therefore, the Mn-doped Ge film constitutes of an almost pure Ge matrix and amorphous $Ge_{1-x}Mn_x$ nanoclusters with high Mn concentration ($x$ = 10−20 %).

Figure. 2 shows MCD spectra of (a) a Ge(001) substrate as a bulk sample (top curve) and (b) the epitaxially grown Mn-doped Ge film with $x_N$ = 6 % (middle curve), measured at 10K with a reflection configuration. A magnetic field of $H$ = 10 kOe was applied perpendicular to the film plane for all the samples. The spectral features of the epitaxially grown Mn-doped Ge film were similar to those of bulk Ge, i.e., the MCD spectrum of the Mn-doped Ge film has peaks at the critical points corresponding to the $E_1$ and $E_0$' transition energies of bulk Ge (hereafter, these energy points of the Mn-doped Ge film are also referred to as $E_1$ and $E_0$'). The MCD intensities of the Mn-doped Ge film at the critical points $E_1$ and $E_0$' seem to be enhanced in comparison with that of bulk Ge, and their magnetic-field dependence showed a clear ferromagnetic hysteresis loop (discussed later). However, a strong broad peak was newly seen at around 2 eV in the Mn-doped Ge film. When the ferromagnetic ordering of FS originates from the intrinsic s,p-d exchange interactions without any ferromagnetic precipitates, the MCD intensity at each critical point and a newly observed MCD peak should have an identical magnetic-field dependence [27]. Curves (a) and (b) in Fig. 3 shows the magnetic-field dependence of the MCD intensities at $E_1$ (= 2.3 eV) and 2.0 eV (the broad peak of Fig. 2(b)), respectively, for the epitaxially grown Mn-doped Ge film with $x_N$ = 6 %, where the data were normalized by their MCD intensities at $H$ = 0 Oe. Although both curves show clear ferromagnetic hysteresis loops, the shape of the hysteresis loop at 2.0 eV (curve (b)) was not identical with that (curve (a)) at the critical point $E_1$. The difference between these hysteresis loops was obtained by subtracting the normalized hysteresis loop at 2.0 eV from that at the critical point $E_1$ (i.e., curve (a)–curve (b)). The result (curve (c) in Fig. 3) shows no ferromagnetic character and showed only the same behavior as the magnetic-filed dependence of bulk Ge. This means that the MCD spectrum of the epitaxially grown Mn-doped Ge film is caused by the overlap of the MCD spectrum of nonmagnetic bulk Ge and ferromagnetic broad MCD peak at around 2.0 eV.

Open and solid circles in Fig. 4 shows the temperature dependence of MCD intensities at (a) $E_1$ (= 2.3 eV) and (b) 2.0 eV of the epitaxially grown Mn-doped Ge film with $x_N$ = 6 %, respectively, where the MCD intensity at $E_1$ plotted in Fig. 4(a) was obtained by subtracting the broad MCD peak shape at around 2.0 eV from the original spectrum. The MCD intensity at 2.0 eV took a large value at low temperatures below 100 K and then the intensity dropped rapidly at around 150 K. The magnetic-field dependence of this MCD intensity (at 2.0 eV) exhibited the hysteretic characteristics up to 100K, indicating that the critical temperature for ferromagnetic ordering was ~100K, which is consistent with the result of magnetization measurements using SQUID. (Note that the MCD intensity observed at over 150K can be attributed to the paramagnetic component.) On the contrary, the deconvoluted MCD intensity at $E_1$ decreased slightly with increasing temperature, which was very similar to the behavior of the MCD intensity at $E_1$ of bulk Ge. The MCD measurements shown in Figs. 2–4 revealed that the Zeeman splitting induced by the s,p-d exchange interactions was not observed at the critical points of the epitaxially grown Mn-doped Ge film, but its ferromagnetic behavior indicates the existence of other ferromagnetic phase. Taking into account the TEM/EDX observations described previously, the ferromagnetic ordering is considered to be caused by the amorphous $Ge_{1-x}Mn_x$ clusters embedded in the Ge matrix.

In order to investigate the magnetism of amorphous $Ge_{1-x}Mn_x$, an amorphous $Ge_{1-x}Mn_x$ film was deposited by MBE on a $SiO_2$ film formed on a Si substrate by thermal oxidation. The growth condition was the same as that of the epitaxially grown Mn-doped Ge film. The Mn concentration of the amorphous $Ge_{1-x}Mn_x$ film was 15 %, which was close to that of the amorphous $Ge_{1-x}Mn_x$ clusters in the epitaxially grown Mn-doped Ge film. *In-situ* RHEED measurements revealed that the deposited film was completely amorphous and no crystalline and poly-crystalline phases were detected during the growth. Fig. 2 (c) shows the MCD spectrum of the amorphous $Ge_{1-x}Mn_x$ film. The spectrum showed no features of the critical points of *single-crystalline* Ge, but a large MCD peak was seen below 2.0 eV (Note that its peak position cannot be determined owing to the detection limit of our MCD system). The MCD spectrum exhibited identical magnetic field dependence at several different energy points, indicating that the ferromagnetic ordering of the amorphous $Ge_{1-x}Mn_x$ film comes from a single phase. Solid triangles in Fig.4 (c) show the temperature dependence of MCD intensity at 1.1 eV for the amorphous $Ge_{1-x}Mn_x$ film with $x$ = 15 %. The temperature dependence is similar to that of the MCD intensity at 2.0 eV for the epitaxially grown Mn-doped Ge film. In addition, it was found that the critical temperature of ferromagnetic ordering of the amorphous $Ge_{1-x}Mn_x$ film (~100K) is close to that of the epitaxial Mn-doped Ge film with amorphous clusters (~100K). Hence, the MCD peak at around 2.0 eV in the epitaxially grown Mn-doped Ge film can be attributed to the amorphous $Ge_{1-x}Mn_x$ clusters, although there exists the peak energy shift for the amorphous $Ge_{1-x}Mn_x$ clusters.  The reason for this energy shift (blue shift) is not clear, but it could be related with the quantum size effect, owing to the nanometer size (~5 nm) of the amorphous $Ge_{1-x}Mn_x$ clusters.

We also found that the critical temperature of ferromagnetic ordering in amorphous $Ge_{1-x}Mn_x$ films ($x$ = 1-15 %) depends on the Mn concentration, and it increases linearly with increasing Mn concentration up to 15 % [28].  The resistivity of the amorphous $Ge_{1-x}Mn_x$ films decreased with increasing Mn concentration and increased with decreasing temperature [28], indicating semiconducting electrical properties.  Furthermore, the amorphous $Ge_{1-x}Mn_x$ films exhibited clear anomalous Hall effect [28].  Therefore, our magnetooptical and magnetotransport measurements indicate that amorphous $Ge_{1-x}Mn_x$ is a homogeneous ferromagnetic semiconductor.  We should emphasize that these results are also consistent with the experimental results previously reported on epitaxially grown Mn-doped Ge films shown by Park *et al* [13].  Amorphous $Ge_{1-x}Mn_x$ nanoclusters and their percolation are considered to be the possible mechanism for the ferromagnetic semiconductor behavior of epitaxially grown Mn-doped Ge films.  This model would be supported by recent investigation using ac susceptibility measurements reported by Li *et al* [20], in which the existence of bound magnetic polarons or clusters and their percolation were observed in epitaxially grown Mn-doped Ge samples that are also free from ferromagnetic intermetallic compounds.

In conclusion, the origin of the ferromagnetism in epitaxially grown Mn-doped Ge films was investigated by crystallographic structure and chemical composition analyses using TEM and EDX and by direct band structure characterization using MCD.  The precipitation of the amorphous ferromagnetic semiconductor nanoclusters of $Ge_{1-x}Mn_x$ is the origin of the ferromagnetism in epitaxially grown Mn-doped Ge films.

Acknowledgments

One of the authors (S.S.) gratefully acknowledges the support of the PRESTO Program of JST,

and a Grant-in-Aid for Scientific Research (16686024) from MEXT. This work was also supported by SORST of JST, Grant-in-Aids for Scientific Research (14076207, 14205003) and IT Program of RR2002 from MEXT.

Figure captions

Fig. 1   Cross-sectional high resolution TEM images of (a) an epitaxially grown Mn-doped Ge film with the nominal Mn content $x_N$ = 6 %, and (b) the same Mn-doped Ge film after the thickness of the TEM specimen was thinned by re-milling.   The scanning areas of the images (a) and (b) are 210 nm and 30 nm in width, respectively.

Fig. 2   MCD spectra of (a) a Ge(001) substrate as a bulk sample, (b) the epitaxially grown Mn-doped Ge film with $x_N$ = 6 %, and (c) a amorphous $Ge_{1-x}Mn_x$ film with the Mn content $x$ = 15 %, measured at 10K with a reflection configuration.   A magnetic field of $H$ = 10 kOe was applied perpendicular to the film plane for all samples.

Fig. 3   Magnetic-field dependence of MCD intensities at (a) $E_1$ (= 2.3 eV) and (b) 2.0 eV (the broad peak of Fig. 2(b)) for the epitaxially grown Mn-doped Ge film with $x_N$ = 6 %, where the data were normalized by their MCD intensities at $H$ = 0 Oe.   The difference between these hysteresis loops was also shown by a loop (c).

Fig. 4   Temperature dependence of MCD intensities at (a) $E_1$ (= 2.3 eV) and (b) 2.0 eV, respectively, of the epitaxially grown Mn-doped Ge film with $x_N$ = 6 %, where the MCD intensity at $E_1$ was obtained by subtracting the broad MCD peak component at around 2.0 eV from the original spectrum.   The temperature dependence of MCD intensity at 1.1 eV for the amorphous $Ge_{1-x}Mn_x$ film with $x$ = 15 % is also shown by solid triangles (curve (c)).

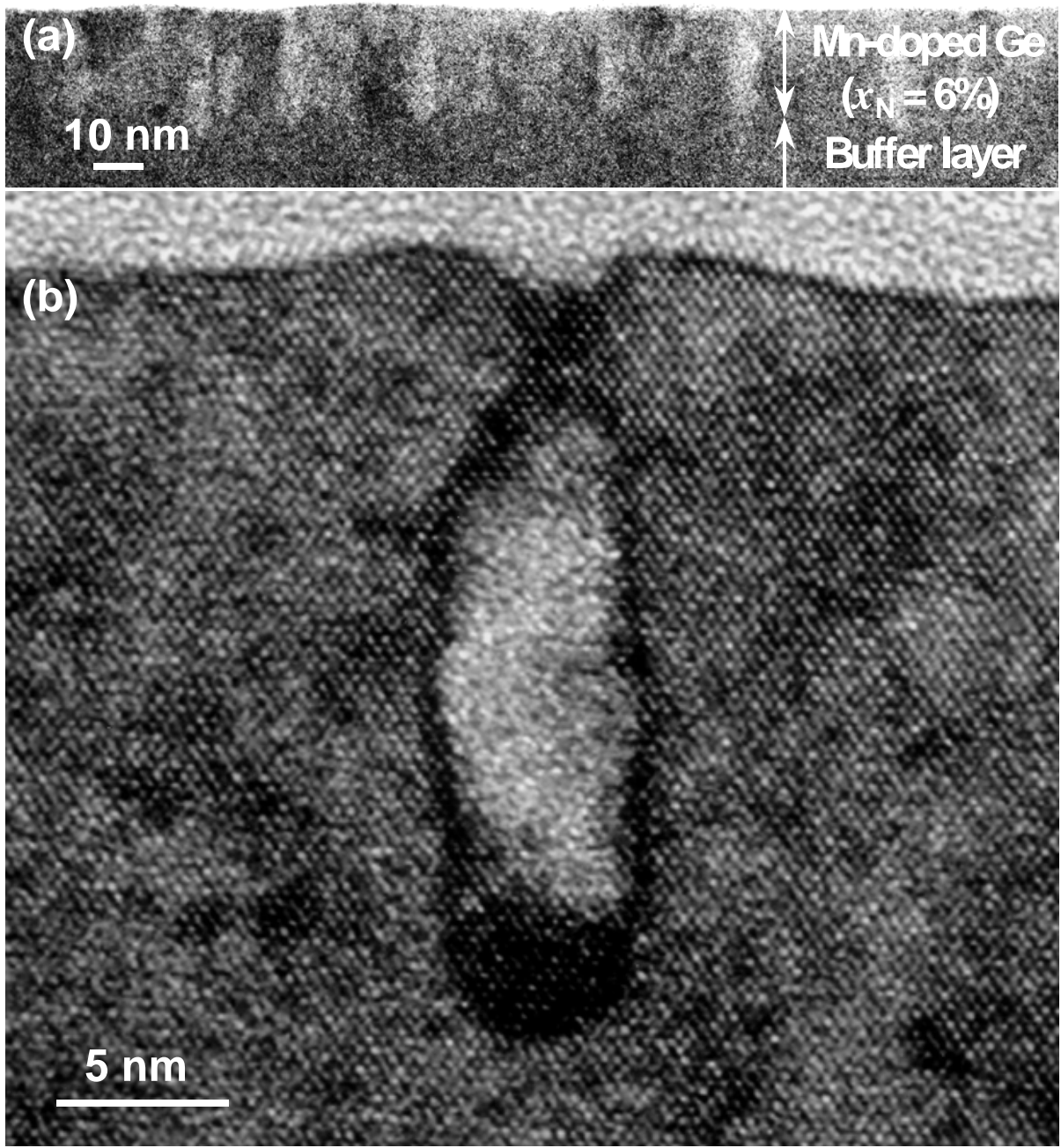

Fig.1 Sugahara et al

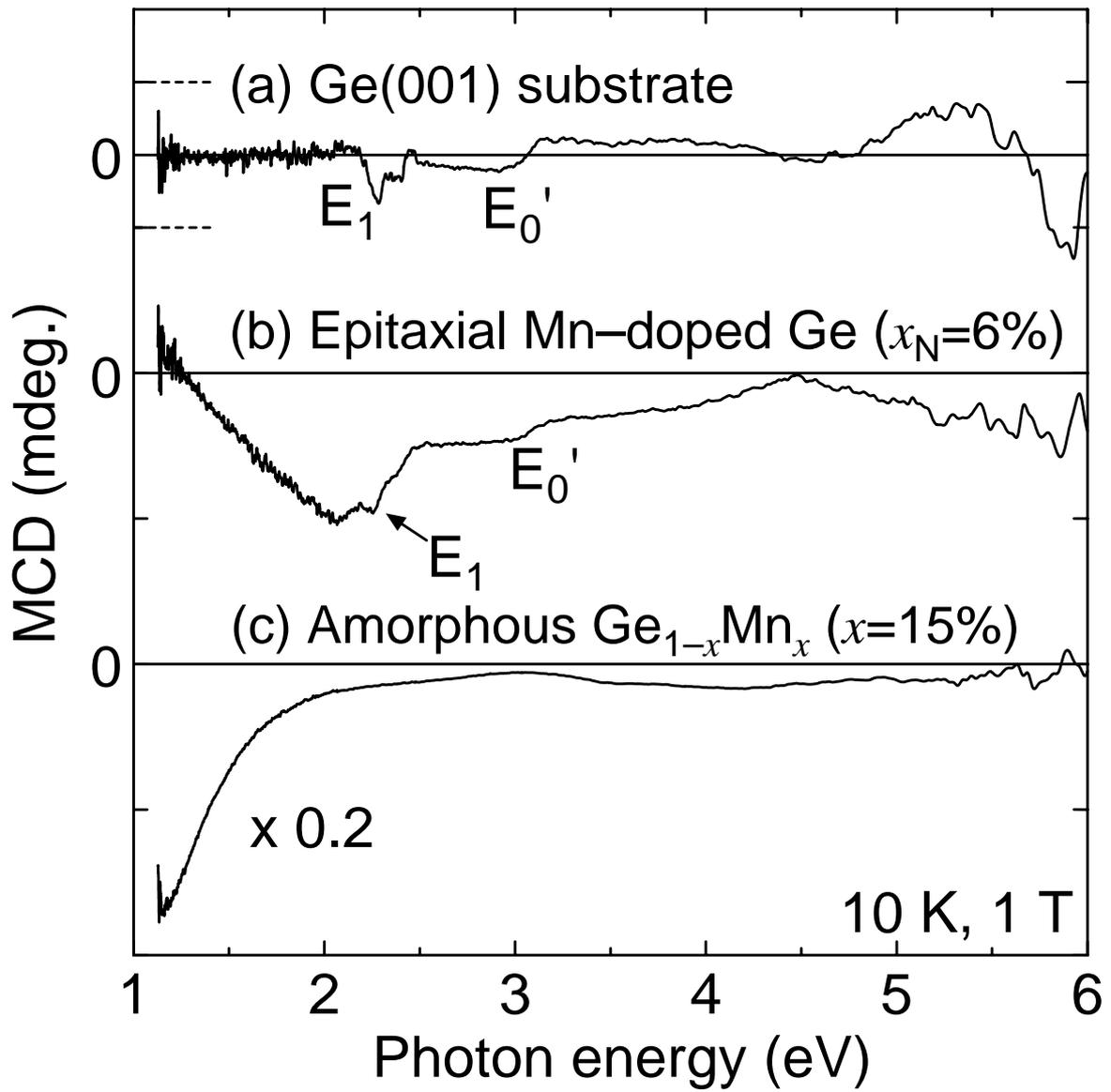

Fig.2 Sugahara et al

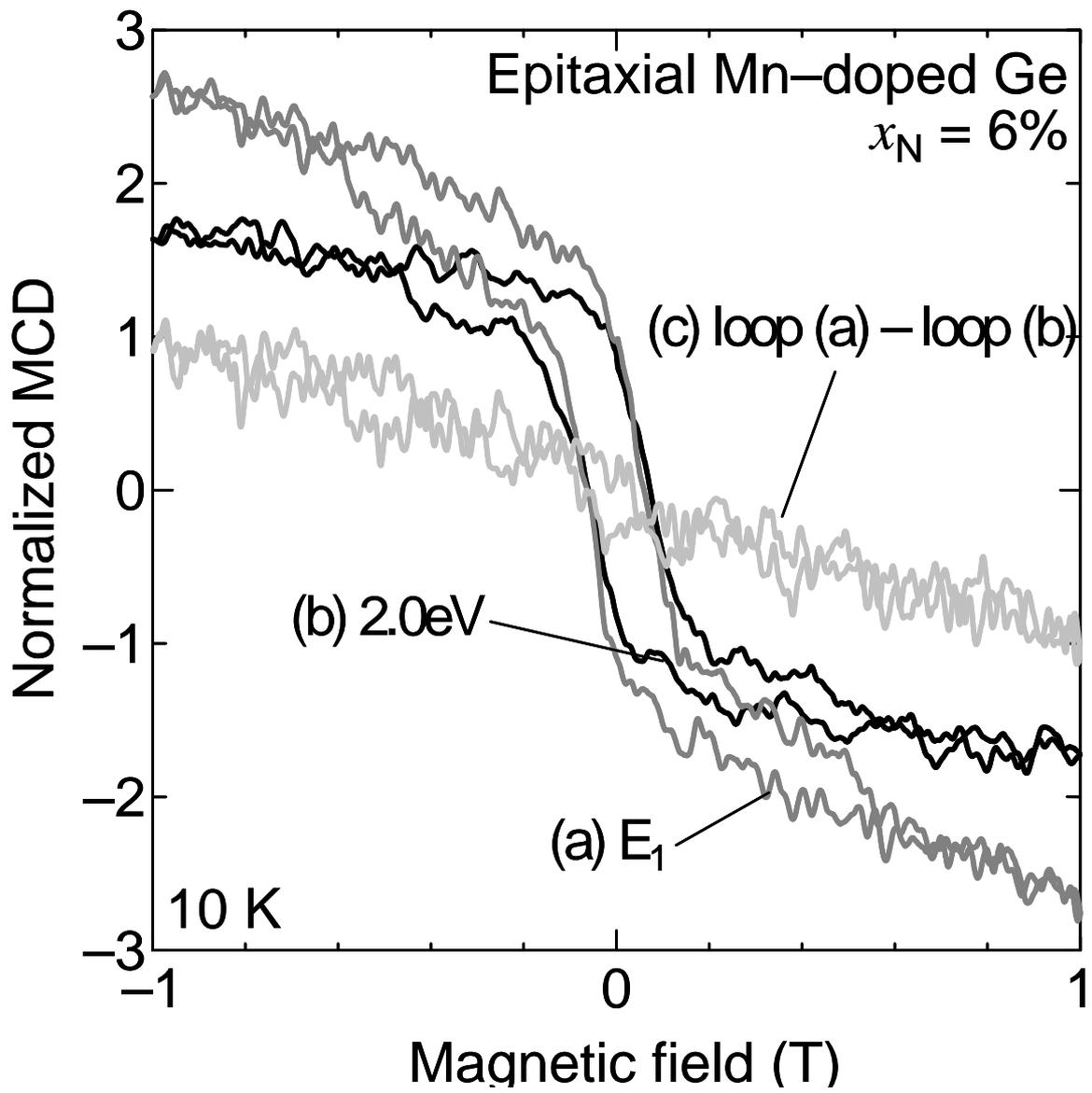

Fig.3 Sugahara et al

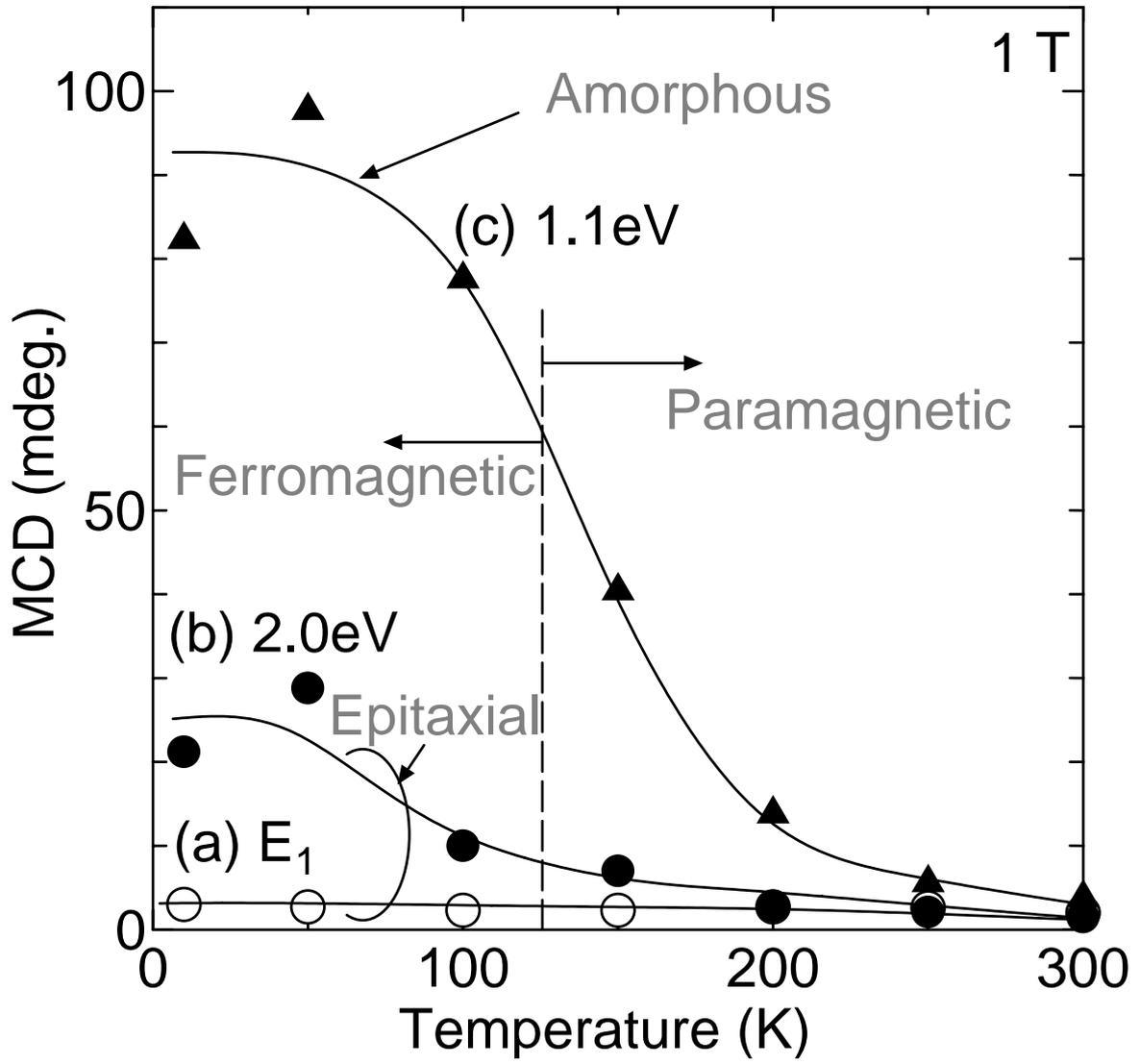

Fig. 4 Sugahara et al